\documentclass[11pt,a4paper]{article}
\usepackage{jcappub}
\usepackage{color}
\usepackage{subfigure}
\input{colordvi.tex}

\newcommand{\simgt}{\lower.5ex\hbox{$\; \buildrel > \over \sim \;$}}
\newcommand{\simlt}{\lower.5ex\hbox{$\; \buildrel < \over \sim \;$}}

\title{
Constraints on general second-order scalar-tensor models
from gravitational Cherenkov radiation
}

\author{Rampei Kimura,}
\author{Kazuhiro Yamamoto}
\emailAdd{rampei@theo.phys.sci.hiroshima-u.ac.jp}
\emailAdd{kazuhiro@hiroshima-u.ac.jp}

\affiliation{
Department of Physical Science, Hiroshima University,
Higashi-Hiroshima 739-8526,~Japan}

\abstract{
We demonstrate that the general second-order scalar-tensor theories, 
which have attracted attention as possible modified gravity models 
to explain the late time cosmic acceleration, could be strongly
constrained from the argument of the gravitational Cherenkov radiation. 
To this end, we consider the purely kinetic coupled gravity and 
the extended galileon model on a cosmological background. 
In these models, the propagation speed of tensor mode could be less 
than the speed of light, which puts very strong constraints from 
the gravitational Cherenkov radiation. 
}


\arxivnumber{1112.4284}

\begin{document}
\maketitle

\section{Introduction}

Cosmological observations 
of type Ia supernovae \cite{Riess,Riess2,Perlmutter}, the cosmic microwave background \cite{Spergel,Komatsu},
and the large scale structures \cite{Percival,sloan,Allen,Rapetti} indicate that the universe undergoes 
a phase of accelerated expansion, and this discovery
opened up a new field in cosmology. 
A number of attempts to explain the origin of the present cosmic acceleration 
have been proposed over the past decade. 
The Einstein's cosmological constant 
might be a possible solution, however, the smallness of value 
of the cosmological constant can not be explained naturally \cite{Weinberg,Weinberg2}.

The possible solution for an accelerated expansion of the universe
at the present time is an alternative theory of gravity. 
So far various modified gravity models have been proposed 
such as the scalar-tensor theories \cite{Amendola,Uzan,Chiba,Bartolo,Perrotta}, 
$f(R)$ gravity \cite{HuSawicki,Starobinsky,Appleby,Nojiri}, 
Dvali-Gabadazde-Porrati (DGP) braneworld model \cite{DGP1,DGP2}, 
and Galileon gravity \cite{galileon,galileon2,galileon3,galileon4,galileon5,galileon6,galileon7,galileon8,KGB,KGB2}. 
In these models, additional degrees of freedom
can mimic the cosmological constant and
lead to cosmic acceleration today.
Most of these theories are a subclass of 
the most general second-order scalar-tensor theory, 
which was first constructed by Horndeski \cite{Horndeski}
and also independently derived by Deffayet {\em et al.} \cite{GenGal}
as an extension of galileon theory.
The most general second-order scalar-tensor theory 
is applied to the late-time accelerated expansion \cite{AKT}
as well as the inflationary models \cite{G-inf,G-inf2,G-inf3,G-inf4}.

The Lagrangian in the most general second-order scalar-tensor theory
contains the coupling of the scalar field $\phi$ 
and its kinetic term $X\equiv -g^{\mu\nu}\nabla_{\mu}\phi\nabla_{\nu}\phi/2$ with gravity,
such as $G_4(\phi,X)R$ and $G_5(\phi,X)G_{\mu\nu}\nabla^{\mu}\nabla^{\nu}\phi$,
where $G_4(\phi,X)$ and $G_5(\phi,X)$ are arbitrary functions of $\phi$ and $X$.
This theory is covariant, but in the presence of a cosmological background
Lorentz invariance could be broken due to these coupling terms.
As a result, the propagation speed of gravitational waves
differs from the speed of light and also
depends on the cosmological background. 

When the propagation speed of gravitational waves is less than the 
speed of light, the gravitons could be emitted through a similar
process to the Cherenkov radiation \cite{Caves,Moore,Moore2}. 
The observation of the high energy cosmic rays puts constraints 
on this process, i.e., the speed of the gravitational waves. 
Assuming a galactic origin for the high energy cosmic rays,
the lower bound on the propagation speed of gravity 
from gravitational Cherenkov radiation is given by \cite{Moore,Moore2}
\begin{eqnarray}
  c-c_T < 2\times10^{-15}c,
  \label{constraint}
\end{eqnarray}
where $c_T$ is the propagation speed of gravity.
When the origin of the high energy cosmic rays is located 
at a cosmological distance, the constraint is four orders of 
magnitude tighter than (\ref{constraint}). 

In the present paper, we show that the argument of 
the gravitational Cherenkov radiation puts a
tight constraint on general second-order scalar-tensor models
on a cosmological background
with a time-varying propagation speed of gravitational waves. 
As a demonstration, we consider two models: 
the purely kinetic coupled gravity \cite{deriCoupling} and 
the extended galileon model \cite{Tsujikawa11}, 
which are a subclass of the most general second-order 
scalar-tensor theory. 

This paper is organized as follows. 
In section \ref{sec:2}, we briefly review the most general second-order 
scalar-tensor theory and the tensor perturbations.
In section \ref{sec:gcr}, we derive the gravitational Cherenkov radiation
on a cosmological background.
In section \ref{sec:3}, we explicitly show that gravitational 
Cherenkov radiation reject the purely kinetic coupled gravity 
model.
In section \ref{sec:4}, we briefly review the extended galileon model 
and see how gravitational Cherenkov radiation 
can tightly constrain model parameters. 
Section \ref{sec:5} is devoted to conclusions.
In appendix \ref{App:Coefficients1}-\ref{App:otherfixedpoint}, we summarize the scalar perturbations, 
derived in \cite{G-inf}, and the coefficients of the scalar 
and tensor perturbations in various regimes 
in the extended galileon model, derived in \cite{Tsujikawa11}.
In appendix \ref{App:negativeab}, a useful constraint on parameters 
in the extended galileon model is demonstrated.

Throughout the paper, we use units in which the speed
of light and the Planck constant are unity, $c=\hbar=1$, 
and $M_{\rm Pl}$ is the reduced Planck mass related 
with Newton's constant by $M_{\rm Pl}=1/\sqrt{8 \pi G}$. 
We follow the metric signature convention $(-,+,+,+)$.

\section{The most general second-order scalar-tensor theory}
\label{sec:2}
The most general second-order scalar-tensor theory 
is described by the action,
\begin{eqnarray}
S=\int d^4x \sqrt{-g} \left(\sum_{i=2}^{5}{\cal L}_{i}+{\cal L}_m\right),
\end{eqnarray}
where 
\begin{eqnarray}
{\cal L}_{2} & = & K(\phi,X),\nonumber\\
{\cal L}_{3} & = & -G_{3}(\phi,X)\Box\phi,\nonumber\\
{\cal L}_{4} & = & G_{4}(\phi,X) R+G_{4,X}[(\Box\phi)^{2}
-(\nabla_{\mu}\nabla_{\nu}\phi)(\nabla^{\mu}\nabla^{\nu}\phi)],\nonumber\\
{\cal L}_{5} & = & G_{5}(\phi,X) G_{\mu\nu}(\nabla^{\mu}\nabla^{\nu}\phi) 
-\frac{1}{6} G_{5,X}[(\Box\phi)^{3}-3(\Box\phi)(\nabla_{\mu}\nabla_{\nu}\phi)\,(\nabla^{\mu}\nabla^{\nu}\phi)\nonumber\\
&&+2(\nabla^{\mu}\nabla_{\alpha}\phi)(\nabla^{\alpha}\nabla_{\beta}\phi)(\nabla^{\beta}\nabla_{\mu}\phi)],
\label{Li}
\end{eqnarray}
where $K,~G_3,~G_4,$ and $G_5$ are arbitrary functions 
of the scalar field $\phi$ and the kinetic term 
$X \equiv -g^{\mu\nu}\nabla_{\mu}\phi\nabla_{\nu}\phi/2$, 
$G_{i\phi}$ and $G_{iX}$ stands for 
$\partial G_i/\partial \phi$ and $\partial G_i/\partial X$,
respectively, and ${\cal L}_m$ is the matter Lagrangian. 
We assume that matter is minimally coupled to gravity.
Note that for the case, $G_4=M_{\rm Pl}^2/2$, 
the Lagrangian ${\cal L}_4$ reproduces the Einstein-Hilbert term.

We consider the tensor perturbations 
in the most general second-order
scalar-tensor theory on a cosmological background, 
and briefly review the results in derived in \cite{G-inf}.
We briefly review the tensor perturbations 
in the most general second-order
scalar-tensor theory, derived in \cite{G-inf}.
The quadratic action for the tensor perturbations can be written as
\begin{eqnarray}
  S_T^{(2)}=\frac{1}{8}\int dtd^3x a^3 
  \left[{\cal G}_T {\dot h}_{ij}^2 -\frac{{\cal F}_T}{a^2}({\vec \nabla}h_{ij})^2\right],
\label{actiongraviton}
\end{eqnarray}
where 
\begin{eqnarray}
{\cal F}_T&\equiv&2\left[G_4
-X\left( \ddot\phi G_{5X}+G_{5\phi}\right)\right],
\label{Ft}\\
{\cal G}_T&\equiv&2\left[G_4-2 XG_{4X}
-X\left(H\dot\phi G_{5X} -G_{5\phi}\right)\right].
\label{Gt}
\end{eqnarray}
Here an overdot denotes differentiation with respect to $t$,
and $H=\dot a/a$ is the hubble parameter.
We find the propagation speed of the tensor perturbations,
\begin{eqnarray}
  c_T^2&\equiv&\frac{{\cal F}_T}{{\cal G}_T}.
  \label{ct2}
\end{eqnarray}
When $G_4=G_4(\phi)$ and $G_5=0$, the propagation speed of 
gravitational waves is equal to the speed of light. 
On the other hand, the propagation 
speed of gravitational waves depends on the cosmological background
in the presence of $G_5$ or $G_4$ being dependent on $X$. 
If the propagation speed of gravitational waves is less than 
the speed of light, it is tightly constrained 
from gravitational Cherenkov radiation.

\def\bfp{{\bf p}}
\def\bfk{{\bf k}}
\def\bfx{{\bf x}}
\def\hath{{\hat h}}
\def\in{{\rm in}}
\section{Gravitational Cherenkov radiation in an expanding universe}
\label{sec:gcr}
In this section, we derive the gravitational Cherenkov radiation
in a cosmological background. For simplicity, we consider a complex scalar
field with the action 
\begin{eqnarray}
  &&S_m=\int d^4x \sqrt{-g}\left[ - 
  g^{\mu\nu}\partial_\mu \Psi^* \partial_\nu\Psi 
  -m^2\Psi^*\Psi-{\xi}R\Psi^*\Psi
  \right].
\label{SM}
\end{eqnarray}
Here we assume the conformal coupling with spacetime 
curvature $\xi=1/6$, for simplicity, but this term can be 
neglected as long as we focus on the subhorizon scales, 
$p/a,m\gg H$, where $p$ is the comoving momentum.
The free part of $\Psi$ can be quantized as 
 \begin{eqnarray}
  &&\hat \Psi(\eta,\bfx)={1\over a}\int{d^3p\over (2\pi)^{3/2}}
  \left[
  \hat b_\bfp \psi_p(\eta) e^{i\bfp\cdot\bfx}
  +\hat c_\bfp^\dagger \psi_p^*(\eta) e^{-i\bfp\cdot\bfx}\right],
\end{eqnarray}
where $\eta$ is the conformal time,
$\hat b_\bfp$ and $\hat c_\bfp^\dagger$ are the annihilation and creation 
operators of the particle and anti-particle, respectively, which satisfy 
the commutation relations $[\hat b_\bfp,\hat b_{\bfp'}^\dagger]
=\delta(\bfp-\bfp')$, $[\hat c_\bfp,\hat c_{\bfp'}^\dagger]=\delta(\bfp-\bfp')$,  
and the mode function obeys
\begin{eqnarray}
  \left({d^2\over d \eta^2}+p^2+m^2 a^2\right)\psi_p(\eta)=0.
\end{eqnarray}
The WKB approximate solution is given by (e.g., \cite{BD})
\begin{eqnarray}
\psi_p(\eta)={1\over \sqrt{2\Omega_p}}
\exp\left[-i\int_{\eta_\in}^\eta \Omega_p(\eta')d\eta'\right]
\end{eqnarray}
with $\Omega_p(\eta)=\sqrt{p^2+m^2a^2}$. 
The WKB approximation is valid for 
\begin{eqnarray}
\Omega_p^2\gg\biggl|
{1\over\Omega_p}{d^2 \Omega_p\over d\eta^2}-{3\over2}
{1\over\Omega_p^2}\left({d\Omega_p\over d\eta}
\right)^2
\biggr|^2,
\end{eqnarray}
which can be satisfied as long as $p/a,m\gg H$. 

On the other hand, the action of the graviton is given by 
eq.~(\ref{actiongraviton}), 
then, we have the quantized graviton field
 \begin{eqnarray}
&&\hath_{\mu\nu}={1\over a}\sqrt{2\over {\cal G}_T}\sum_\lambda
\int{d^3k\over (2\pi)^{3/2}}
  \biggl[
  \varepsilon_{\mu\nu}^{(\lambda)}\hat a_\bfk h_k(\eta) e^{i\bfk\cdot\bfx}
+\varepsilon_{\mu\nu}^{(\lambda)}{\hat a_\bfk}^\dagger h_k^{*}(\eta) 
e^{-i\bfk\cdot\bfx}\biggr],
 \end{eqnarray}
where $\varepsilon_{\mu\nu}^{(\lambda)}$ is the polarization tensor, 
$\hat a_\bfk^\dagger$ and $\hat a_\bfk$ are the creation and 
annihilation operators, which satisfy the commutation relation 
$[\hat a_\bfk,\hat a_{\bfk'}^\dagger]=\delta(\bfk-\bfk')$, 
and the mode function satisfies
\begin{eqnarray}
  \left({d^2\over d \eta^2}+c_s^2k^2-{a''\over a}\right)h_k (\eta)=0.
\end{eqnarray}
For the case $c_s \sim {\cal O}(1)$ and $c_sk/a\gg H$, we may write
\begin{eqnarray}
  h_k(\eta)={1\over\sqrt{2\omega_k}}\exp\left[-i\int_{{\eta}_\in}^\eta\omega_k(\eta')d\eta'\right],
\end{eqnarray}
where we defined $\omega_k=c_sk$, and the approximate solution is valid as long as $c_sk/a\gg H$.
The interaction part of the action (\ref{SM}) is given by
\begin{eqnarray}
  S_I&=&-\int dtd^3x a h_{ij}\partial_i\Psi\partial_j\Psi^*
\nonumber
\\
 &=&- \int d\eta d^3x h_{ij}\partial_i\psi\partial_j\psi^*,
\end{eqnarray}
where we defined $\psi=a\Psi$, and the interaction Hamiltonian is 
\begin{eqnarray}
  H_I&=& a\int d^3x  h_{ij}\partial_i\Psi\partial_j\Psi^*.
\end{eqnarray}

\begin{figure}[t]
 \begin{center}
  \includegraphics[width=70mm]{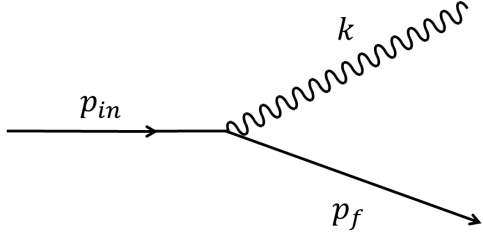}
 \end{center}
 \caption{Feynman diagram for the process}
 \label{fig:one}
\end{figure}

In order to evaluate the gravitational Cherenkov radiation, we adopt the
method developed in \cite{KNY,YN}.
Based on the in-in formalism \cite{Wein}, the lowest order contribution is
given by
\begin{eqnarray}
&&\left<Q(t)\right>=
{i^2} 
\int^t_{t_\in}dt_2
\int^{t_2}_{t_\in}dt_1\left<[H_I(t_1),[H_I(t_2),Q]]\right>.
\end{eqnarray}
We consider the expectation value of the number operator and the initial 
state with the one particle state with the initial momentum, i.e., 
${\hat b}^\dagger_{\bfp_\in}|0\rangle$. Then the lowest-order contribution of the process
so that one graviton with the momentum $\bfk$ is emitted from the massive particle with
the initial momentum $\bfp_\in$, as shown in fig.~\ref{fig:one},
is written as \cite{AEL}
\begin{eqnarray}
&&\left<{\hat a}^{\dagger(\lambda)}_\bfk{\hat a}^{(\lambda)}_\bfk \right>=
2\Re\int^t_{t_\in}dt_2\int^{t_2}_{t_\in}dt_1
\left<H_I(t_1)
{\hat a}^{\dagger(\lambda)}_\bfk{\hat a}^{(\lambda)}_\bfk 
H_I(t_2)\right>.
\end{eqnarray}
Then, the total radiation energy from the scalar particle can be estimated as 
$E=\sum_\lambda\sum_\bfk (\omega_k/ a)$ $\bigl<{\hat a}^{\dagger(\lambda)}_\bfk
{\hat a}^{(\lambda)}_\bfk \bigr>$,
which leads to
\begin{eqnarray}
&&E=\sum_\lambda \int{d^3k\over (2\pi)^3}{\omega_k\over a}
\biggl|
\int_{\eta_\in}^\eta d\eta_1
{1\over a(\eta_1)} 
\sqrt{{2\over {\cal G}_T}}
h_k(\eta_1)\psi_{\bfp_f}(\eta_1)
\psi_{\bfp_\in}^*(\eta_1)\epsilon_{ij}p_{\rm in}^i p^j_f
\biggr|^2,
\end{eqnarray}
where $\bfp_f+\bfk=\bfp_\in~ (p_f^i+k^i=p_\in^i)$. With the use of the relation 
$\sum_\lambda \bigl|\epsilon_{ij}p_\in^i p_f^j\bigr|^2=p_\in^4\sin^4\theta$,
we have
\begin{eqnarray}
E&=&\int{d^3k\over (2\pi)^3}{\omega_k\over a}p_\in^4\sin^4\theta
\biggl|
\int_{\eta_\in}^\eta d\eta_1
{1\over a(\eta_1)} 
\sqrt{{2\over {\cal G}_T}}
h_k(\eta_1)\psi_{\bfp_f}(\eta_1)
\psi_{\bfp_\in}^*(\eta_1)
\biggr|^2.
\nonumber
\\
\label{EA}
\end{eqnarray}
We are now interested in the subhorizon scales, 
$k/a,~p/a,~m,~c_sk/a \gg H$, and 
the situation so that the scale factor $a$ 
is constant, then we can approximate as 
\begin{eqnarray}
&&\int_{\eta_\in}^\eta d\eta_1
{1\over a(\eta_1)} 
\sqrt{{2\over {\cal G}_T}}
h_k(\eta_1)\psi_{\bfp_f}(\eta_1)
\psi_{\bfp_\in}^*(\eta_1)
\nonumber
\\
 &&~~~~~~
\simeq{1\over a}\sqrt{{2\over {\cal G}_T}}
    {1\over \sqrt{2\omega_k}}{1\over \sqrt{2\Omega_{\bfp_\in}}}{1\over \sqrt{2\Omega_{\bfp_f}}}
\int_{\eta_\in}^\eta d\eta_1
    \exp\left[i(\Omega_\in-\Omega_f-\omega_k)(\eta_1-\eta_{\rm ini})    \right].
\end{eqnarray}
Then the total radiation energy eq.~(\ref{EA}) reduces to
\begin{eqnarray}
E&\simeq&{1\over 4{\cal G}_Ta^3}\int{d^3k\over (2\pi)^3}{p_\in^4\sin^4\theta\over \Omega_f\Omega_\in}
\frac{2\pi T}{a} \delta(\Omega_\in-\Omega_f-\omega_k),
\end{eqnarray}
Here we assumed the long time duration of the integration, 
\begin{eqnarray}
&&\biggl|
\int_{\eta_\in}^\eta d\eta_1
    \exp\left[i(\Omega_\in-\Omega_f-\omega_k)(\eta_1-\eta_{\rm ini})\right]
\biggr|^2
\simeq \frac{2\pi T}{a} \delta(\Omega_\in-\Omega_f-\omega_k),
\end{eqnarray}
where $T/a=\eta-\eta_\in$. 
Then, we have the expression in the relativistic limit of the massive particle, 
$p_\in/a\gg m$,
\begin{eqnarray}
{dE\over dt}&=&{p_\in^2\over 4{\cal G}_Ta^4}\int_0^\infty{dkk^2\over 2\pi}
\int_{-1}^1d(\cos\theta){\sin^4\theta}
\delta(\Omega_\in-\Omega_f-\omega_k).
\end{eqnarray}
Now consider the delta-function, which can be written as
\begin{eqnarray}
\delta(\Omega_\in-\Omega_f-\omega_k)=2\Omega_f\delta(\Omega_f^2-(\Omega_\in-\omega_k)^2)
\end{eqnarray}
where
$\omega_k=c_sk$,
$\Omega_\in=\sqrt{\bfp^2_\in+a^2m^2}$, and 
$\Omega_f=\sqrt{(\bfp_\in-\bfk)^2+a^2m^2}$.
With the use of the fact
\begin{eqnarray}
&&\Omega_f^2-(\Omega_\in-\omega_k)^2
=-2p_\in k\left(
\cos\theta-{c_s\over\beta}-{(1-c_s^2)k\over 2p_\in}\right),
\end{eqnarray}
where we defined $\beta=p_\in/\sqrt{p_\in^2+m^2a^2}$ and $p_\in^2=|\bfp_\in|^2$, 
we find (cf. eq.(3.2) in reference by Moore and Nelson \cite{Moore})
\begin{eqnarray}
{dE\over dt}&=&{p_\in^2\over 4{\cal G}_Ta^4}\int_0^{k_{\rm max}}{dkk\over 2\pi}{\sin^4\theta}
\end{eqnarray}
with
\begin{eqnarray}
&&\cos\theta={c_s\over\beta}+{(1-c_s^2)k\over 2p_\in}
\end{eqnarray}
and
\begin{eqnarray}
&&k_{\rm max}={2p_\in\over 1-c_s^2}\left(1-{c_s\over \beta}\right).
\end{eqnarray}
Assuming $\beta\sim 1$, we have $k_{\rm max}\simeq 2p_\in/(1+c_s)$ and
\begin{eqnarray}
{dE\over dt}\simeq{p_\in^2\over 8\pi {\cal G}_T a^4}4(1-c_s)^2\int_0^{k_{\rm max}} dk k 
\left(1-{k\over {k_{\rm max}}}\right)^2,
\end{eqnarray}
which yields (cf.\cite{Moore})
\begin{eqnarray}
{dE\over dt}\simeq{G_N p_\in^4\over a^4}{4(1-c_s)^2\over3(1+c_s)^2},
\end{eqnarray}
where we introduce the Newtonian gravity constant by $G_N=1/16\pi{\cal G}_T$.
One may notice that this definition of the Newton's constant 
is slightly different from that in the most general second-order scalar-tensor theory
(cf. \cite{Vainstein2nd}), however, it does not affect the constraints significantly.
Our results are consistent with those in Ref.~\cite{Moore}.
Then, a particle with momentum $p$ cannot possibly have been traveling for 
longer than 
\begin{eqnarray}
t_{\rm}\sim{a^4\over G_N}{(1+c_s)^2\over 4(1-c_s)^2}{1\over p^3}.
\end{eqnarray}
Therefore, the highest energy cosmic ray put the constraint on the sound speed of
the graviton 
\begin{eqnarray}
{(1-c_s)}\simlt 2\times10^{-17}\left({10^{11}{\rm GeV}\over p}\right)^{3/2} 
\left({1 {\rm Mpc}\over ct}\right)^{1/2}.
\label{result}
\end{eqnarray}

Since we are considering the theory on a cosmological background, the sound speed of 
the graviton is determined by the cosmological evolution of the background field. 
This situation is slightly different from that in ref.~\cite{Moore}. 
However, as we have shown in this section, the theory on a cosmological background 
can be constrained from the gravitational Cherenkov radiation when the speed of the 
graviton is smaller than that of light. 
Also, there are no the higher order nonlinear interaction terms of the graviton like 
the galileon cubic term that becomes important at short distance \cite{Gao}, which suggests 
that the nonlinear interactions of the gravitons can be ignored.

\section{Purely kinetic coupled gravity}
\label{sec:3}
We first consider the modified gravity model, 
whose action contains a nonminimal derivative coupling to gravity.
The action proposed by Gubitosi and Linder \cite{deriCoupling} is given by
 \begin{eqnarray}
   S=\int d^4x \sqrt{-g}\left[\frac{M_{\rm Pl}^2}{2}R+X
     +\frac{\lambda}{M_{\rm Pl}^2}G^{\mu\nu}\nabla_{\mu}\phi\nabla_{\nu}\phi\right],
 \end{eqnarray}
where $\lambda$ is a dimensionless constant.
In this model, the arbitrary functions in eq.~(\ref{Li}) correspond to 
$K=X$, $G_3=0$, $G_4=M_{\rm Pl}^2/2$, and $G_5=-\lambda\phi/M_{\rm Pl}^2$.
Using the matter density parameter $\Omega_m=\rho_m/3M_{\rm Pl}^2H^2$,
the modified Friedmann equation can be written as
$1=\Omega_m+\Omega_{\phi}$, where 
 \begin{eqnarray}
\Omega_{\phi}=\frac{X}{3M_{\rm Pl}^2H^2}\left(1+18C\right).
 \end{eqnarray}
Here we defined the key parameter,
 \begin{eqnarray}
C\equiv\frac{\lambda H^2}{M_{\rm Pl}^2} > C_{*},
\label{C_inequality}
 \end{eqnarray}
where $C_*=-1/18$.
The second inequality is the condition which ensures
the positivity of the energy density of the scalar field, $\Omega_{\phi} > 0$.
Using the gravity equations and 
the energy density $\rho_{\phi}$ and the pressure $p_{\phi}$ for the scalar field,
the effective equation of state, $w_{\rm eff}\equiv p_{\phi}/\rho_{\phi}$,
can be written as
 \begin{eqnarray}
   w_{\rm eff}=\frac{1+30C}{1+(24-6\Omega_{\phi})C+108(1+\Omega_{\phi})C^2}.
 \end{eqnarray}
Gubitosi and Linder showed that if the deviation parameter at the present time, 
$\delta \equiv (C_*-C)/C_*|_{z=0}$, satisfies $\delta<2/5$,
corresponding to the condition for negative pressure $w_{\rm eff}<0$, 
the kinetic term $X$ behaves as the cosmological constant
around the present time.

The propagation speed of gravitational waves (\ref{ct2}) can be written as
 \begin{eqnarray}
   c_T^2&=&\frac{M_{\rm Pl}^2+2\lambda X/M_{\rm Pl}^2}{M_{\rm Pl}^2-2\lambda X/M_{\rm Pl}^2}.
 \label{GW_PKCG}
 \end{eqnarray}
The condition for avoiding ghosts of the tensor perturbations,
${\cal G}_T > 0$, 
is $\delta > \Omega_{\phi}(\Omega_{\phi}-3)$,
which is automatically satisfied,
while the condition for avoiding instability $c_T^2 \geq 0$ is
 \begin{eqnarray}
   \delta \geq \frac{\Omega_{\phi}}{\Omega_{\phi}+3}.
 \end{eqnarray}
Therefore, $\delta > 0$ is required for avoiding ghost-instability.
Thus the theoretically allowed parameter range is
 \begin{eqnarray}
   0 < \delta < {2 \over 5},
 \label{condition:PKCGa}
 \end{eqnarray}
which is equivalent with
\begin{eqnarray}
 -{1 \over 18} < C(z=0) < -{1 \over 30}.
 \label{condition:PKCGb}
 \end{eqnarray}

The propagation speed of gravitational waves in terms of $\Omega_{\phi}$
is rephrased as 
 \begin{eqnarray}
   c_T^2 &=&\frac{(3+\Omega_{\phi})\delta-\Omega_{\phi}}
   {(3-\Omega_{\phi})\delta+\Omega_{\phi}}.
   \label{ct2:pkcg}
 \end{eqnarray}
The constraints from gravitational Cherenkov radiation 
$c_T > 1- \epsilon$, where $\epsilon=2\times10^{-15}$, 
reads $\delta > 1-{\cal O}(\epsilon)$ from eq.~(\ref{ct2:pkcg}), 
which contradicts with the condition (\ref{condition:PKCGa}). 
Equivalently, from eqs.~(\ref{C_inequality}) and (\ref{condition:PKCGb}),
$\lambda$ is always negative, therefore, the propagation speed 
of gravitational waves is always smaller than unity from 
eq.~(\ref{GW_PKCG}). 
Thus this purely kinetic coupled gravity is inconsistent 
with the constraint from the gravitational Cherenkov radiation for 
any theoretically allowed parameter $\lambda$.

\section{Extended galileon model}
\label{sec:4}
In this section, we consider the model proposed 
by De Felice and Tsujikawa \cite{Tsujikawa11},
which is an extension of the covariant galileon model \cite{CovGalileon}.
In this model, the arbitrary functions has the following form, 
\begin{eqnarray}
K&=&-c_{2}M_{2}^{4(1-p_{2})}X^{p_{2}},\nonumber\\
G_{3}&=&c_{3}M_{3}^{1-4p_{3}}X^{p_{3}},\nonumber\\
G_{4}&=&\frac{1}{2}M_{{\rm pl}}^{2}-c_{4}M_{4}^{2-4p_{4}}X^{p_{4}},\nonumber\\
G_{5}&=&3c_{5}M_{5}^{-(1+4p_{5})}X^{p_{5}},
\label{GCGM}
\end{eqnarray}
where $c_i$ and $p_i$ are the model parameters and 
$M_i$ are constants with dimensions of mass.
We impose the conditions that the tracker solution is characterized by
$H \dot{\phi}^{2q}={\rm const}$ 
and the energy density of the scalar field 
is proportional to $\dot{\phi}^{2p}$. 
These conditions
enable us to reduce the model parameters,
which is given by $p_{2}=p$, $p_{3}=p+(2q-1)/2$,
$p_{4}=p+2q$, and $p_{5}=p+(6q-1)/2$
\footnote{Kimura and Yamamoto considered the case :
$p=1$, $q=n-1/2$, $c_4=0$, and $c_5=0$ \cite{KY1}.}.
Note that the covariant Galileon model corresponds to $p=1$ and $q=1/2$.

\subsection{Cosmological Dynamics}
In this subsection, we briefly review the background dynamics 
in the extended galileon model.
For convenience, we write the mass dimension constants as
\begin{eqnarray}
M_{2}&\equiv&(H_{{\rm dS}}M_{{\rm Pl}})^{1/2},\nonumber\\
M_{3}&\equiv&\left(\frac{{M_{{\rm Pl}}}^{1-2p_{3}}}{{H_{{\rm dS}}}^{2p_{3}}}\right)^{1/(1-4p_{3})},\nonumber\\
M_{4}&\equiv&\left(\frac{{M_{{\rm Pl}}}^{2-2p_{4}}}{{H_{{\rm dS}}}^{2p_{4}}}\right)^{1/(2-4p_{4})},\nonumber\\
M_{5}&\equiv&\left(\frac{{H_{{\rm dS}}}^{2+2p_{5}}}{{M_{{\rm Pl}}}^{1-2p_{5}}}\right)^{1/(1+4p_{5})}\,,
\end{eqnarray}
where $H_{\rm dS}$ is the hubble parameter at the de-Sitter point.
At the de Sitter point $\dot{H}=0$ and $\ddot{\phi}=0$, 
we obtain the following relations from the gravitational and scalar field equations
\begin{eqnarray}
c_{2}&=&\frac{3(3\alpha-4\beta+2)}{2}\left(\frac{2}{x_{{\rm dS}}^{2}}\right)^{p},\nonumber\\
c_{3}&=&\frac{\sqrt{2}\left[3(p+q)(\alpha-\beta)+p\right]}{2p+q-1}\left(\frac{2}{x_{{\rm dS}}^{2}}\right)^{p+q},
\end{eqnarray}
where $x\equiv \dot{\phi}/HM_{{\rm pl}}$ and  
\begin{eqnarray}
\alpha&\equiv&\frac{4(2p_{4}-1)}{3}\left(\frac{x_{{\rm dS}}^{2}}{2}\right)^{p_{4}}c_{4},\nonumber\\
\beta&\equiv&2\sqrt{2}\, p_{5}\left(\frac{x_{{\rm dS}}^{2}}{2}\right)^{p_{5}+1/2}c_{5}.
\end{eqnarray}
Thus this model is characterized by only four parameters $p$, $q$, $\alpha$, and $\beta$.
In order to simplify the analysis, we introduce the following variables, 
\begin{eqnarray}
r_{1}&\equiv&\left(\frac{x_{{\rm dS}}}{x}\right)^{2q}\left(\frac{H_{{\rm dS}}}{H}\right)^{1+2q},\nonumber\\
r_{2}&\equiv&\left[\left(\frac{x}{x_{{\rm dS}}}\right)^{2}\frac{1}{r_{1}^{3}}\right]^{\frac{p+2q}{1+2q}},
\end{eqnarray}
and the radiation density parameter $\Omega_{r}\equiv \rho_{r}/3H^{2}M_{{\rm pl}}^{2}$.
Note that the de Sitter fixed point corresponds to $(r_{1},r_{2},\Omega_{r})=(1,1,0)$.

Along the tracker $r_1=1$, 
the evolution of $r_{2}$ and $\Omega_{r}$ are governed by 
the following differential equations,
\begin{eqnarray}
r_{2}' & = & \frac{(1+s)(\Omega_{r}+3-3r_{2})}{sr_{2}+1}\, r_{2}\,,\\
\Omega_{r}' & = & \frac{\Omega_{r}-1-3r_{2}-4sr_{2}}{sr_{2}+1}\,\Omega_{r}\,,
\end{eqnarray}
where a prime denotes a derivative with respect to $N=\ln a$
and only one parameter $s=p/2q$ determines 
the background dynamics in the case of the tracker solution.
In this case, the density parameter of the scalar field is 
simply given by $\Omega_{\phi}=r_2$,
satisfying the constraint $1=\Omega_{\phi}+\Omega_m+\Omega_r$.
Integrating these equations yields the following
algebraic equations, 
\begin{eqnarray}
r_{2} & = & b_{1}a^{4(1+s)}{\Omega_{r}}^{1+s}\,,\\ 
b_{1}a^{4(1+s)}{\Omega_{r}}^{1+s} & = & 1-\Omega_{r}(1-b_{2}a)\,,
\label{eom_omegar}
\end{eqnarray}
where the integration constants are given by
\begin{equation}
  b_{1}=\frac{1-\Omega_{m0}-\Omega_{r0}}{\Omega_{r0}^{1+s}}\,,
  \qquad b_{2}=-\frac{\Omega_{m0}}{\Omega_{r0}},
\end{equation}
and $\Omega_{m0}$ and $\Omega_{r0}$ are the 
matter and radiation density parameter at present, respectively.
To see how the Friedmann equation is modified, 
we rewrite the algebraic equation (\ref{eom_omegar}) 
in terms of the hubble parameter $H$, then we find
\begin{eqnarray}
  \left({H \over H_0}\right)^2&=&(1-\Omega_{m0}-\Omega_{r0})\left({H \over H_0}\right)^{-2s}
  +\Omega_{m0}a^{-3}+\Omega_{r0}a^{-4}.
\label{DvaliTurnerEq}
\end{eqnarray}
This modified Friedmann equation
is known as the Dvali-Turner model \cite{DvaliTurner}. 
The authors in \cite{KY1} placed the observational constraints 
on this modified Friedmann equation (\ref{DvaliTurnerEq}) 
in the special case $p=1$
using type Ia supernovae and the CMB shift parameter 
and showed that the model parameter $s$ has to be small, $s \ll 1$,
in order to be consistent with cosmological observations\footnote{
Observational constraints on eq.~(\ref{DvaliTurnerEq})
from type Ia supernovae, cosmic microwave background, and 
baryon acoustic oscillations  
including the cosmic curvature $K$ in the context of 
the extended galileon model has been recently studied 
by De Felice and Tsujikawa \cite{ObsExtGalileon}.
They found that the parameter $s$ is constrained to be 
$s =0.034_{-0.034}^{+0.327}~(95\%~{\rm CL})$ 
in the flat case $K=0$.
}.

\subsection{Conditions}

In this subsection, we summarize the theoretically allowed 
parameter space in the extended galileon model, 
discussed in \cite{Tsujikawa11}, 
and show that the constraint from gravitational Cherenkov 
radiation is crucial.
To avoid ghost-instabilities, we must impose the conditions,
${\cal G}_T>0$, $c_T^2>0$, ${\cal G}_S>0$, and $c_S^2>0$ 
in the history of the universe.
The coefficients in the tensor and scalar perturbation equations
in terms of $r_1$, $r_2$, $\Omega_r$, and the model parameters
are listed in appendix \ref{App:Coefficients2}. 
We find that the propagation speed of gravitational waves 
along the tracker $r_1=1$ is written
 \begin{eqnarray}
  c_{T}^2=\frac{2(1-2p-4q)(2q+pr_2)+3\alpha (2q+pr_2) r_2-3\beta (1-2p-4q) 
    (3-3r_2+\Omega_r)r_2}{(1-2p-4q)[2+3(\alpha-2\beta)r_2](2q+pr_2)}.\nonumber\\
 \label{ct2tracker}
 \end{eqnarray}
Note that eq.~(\ref{ct2tracker}) reduces $c_T^2=1$ 
when $\alpha=\beta=0$, which correspond to $G_4=M_{\rm Pl}^2/2$ and $G_5=0$.
We further impose no-instability condition at $r_2=r_{2,{\rm min}}$,
where a minimum of propagation speed of gravitational waves $c_T^2$ is 
located.
Setting $r_1=1$ and $\Omega_r\simeq 0$, 
the minimum of $c_T^2$ is given by eq.(\ref{ct2tracker})
at $r_2=r_{2,min}$,
 \begin{eqnarray}
   r_{2,min}&=&\biggl[2(3+2p)(1-2p-4q)q\,\beta-8p\,q(p+2q)\alpha
     \pm \sqrt{3\,\Gamma_1}\biggr]/\Gamma_2,
 \end{eqnarray}
where 
 \begin{eqnarray}
   \Gamma_1&=&(1-2p-4q)(p+2q)q \,\beta 
   \times[4(p+2q)(p-3q\,\alpha)\alpha\nonumber\\
   &&+2(1-2p-4q)\{(3+2p)-3(3-4q)\beta\}\beta
   +3[3-16q(1-2q)-2p(3-8q)]\alpha\,\beta],\nonumber\\
   \Gamma_2&=&4p^2(p+2q)\alpha-18(p+2q)(1-2p-4q)\beta^2+(1-2p-4q)[2p(3+2p)+9(p+2q)\alpha]\beta.\nonumber\\
 \end{eqnarray}
The conditions for avoiding ghost-instabilities 
in the regimes along the tracker are given by
\begin{eqnarray}
  {\cal G}_S|_{r_1=1, r_2 \ll 1} &>& 0\,, \qquad  {\cal G}_S|_{\rm de~Sitter} > 0\,,\nonumber\\
  c_S^2|_{r_1=1, r_2 \ll 1} &\geq& 0\,, \qquad  ~c_S^2|_{\rm de~Sitter} \geq 0\,,\nonumber\\
  {\cal G}_T|_{r_1=1, r_2 \ll 1} &>& 0\,, \qquad  {\cal G}_T|_{\rm de~Sitter} > 0\,,\nonumber\\
  c_T^2|_{r_1=1, r_2 \ll 1} &\geq& 0\,, \qquad  ~ c_T^2|_{\rm de~Sitter} \geq 0\,,\nonumber\\
  c_T^2|_{r_2,{\rm min}} &>& 0.
\label{condition:1}
\end{eqnarray}
If the initial condition of $r_1$ is $r_1 \ll 1$, 
we then must impose the conditions for avoiding ghost-instabilities
in the regime $r_1 \ll 1$ and $r_2 \ll 1$, which is given by
\begin{eqnarray}
  {\cal G}_S|_{r_1 \ll 1, r_2 \ll 1} &>& 0\,, \qquad  c_S^2|_{r_1 \ll 1, r_2 \ll 1} \geq 0, \nonumber\\
  {\cal G}_T|_{r_1 \ll 1, r_2 \ll 1} &>& 0\,, \qquad  c_T^2|_{r_1 \ll 1, r_2 \ll 1} \geq 0.
\label{condition:2}
\end{eqnarray}
We also impose the condition that the other fixed points $r_a$ and $r_b$ 
(see appendix \ref{App:otherfixedpoint}) is not real or outside 
the interval $0<r_1 \leq 1$, which is given by
\begin{eqnarray}
  \Delta<0~~~~{\rm or}~~~~
  r_{a,b} < 0 ~~~~{\rm or}~~~~ r_{a,b} \geq 1.
\label{condition:3}
\end{eqnarray}
Note that as long as the initial condition of $r_1$ is near
$r_1=1$ and the scalar field follows the tracker from early stage, 
these conditions (\ref{condition:2}) and (\ref{condition:3}) do 
not have to be imposed.

Let us classify the constraints into four classes: 
(a) the constraint from the gravitational 
Cherenkov radiation, which is given by eq.~(\ref{constraint}) 
and eq.~(\ref{ct2tracker}) with setting $c_T=c_T|_{z=0}$,
(b) the theoretical constraint (\ref{condition:1}) to avoid the ghost-instabilities
when the scalar field follows the tracker solution from early stage,
assuming that the tracker is near $r_1=1$ initially,
(c) the theoretical constraint (\ref{condition:2}) and (\ref{condition:3}) 
in addition to (\ref{condition:1}) to avoid the ghost-instabilities
when the scalar field does not follow the tracker solution initially, 
assuming that the initial condition of $r_1$ is sufficiently small,
(d) the other constraint from the cosmological observations, 
type Ia supernovae, the shift parameter from the cosmic microwave background,
and the baryon acoustic oscillations.

\begin{figure}[t]
  \begin{tabular}{cc}
   \begin{minipage}{0.5\textwidth}
    \begin{center}
     \includegraphics[scale=0.4]{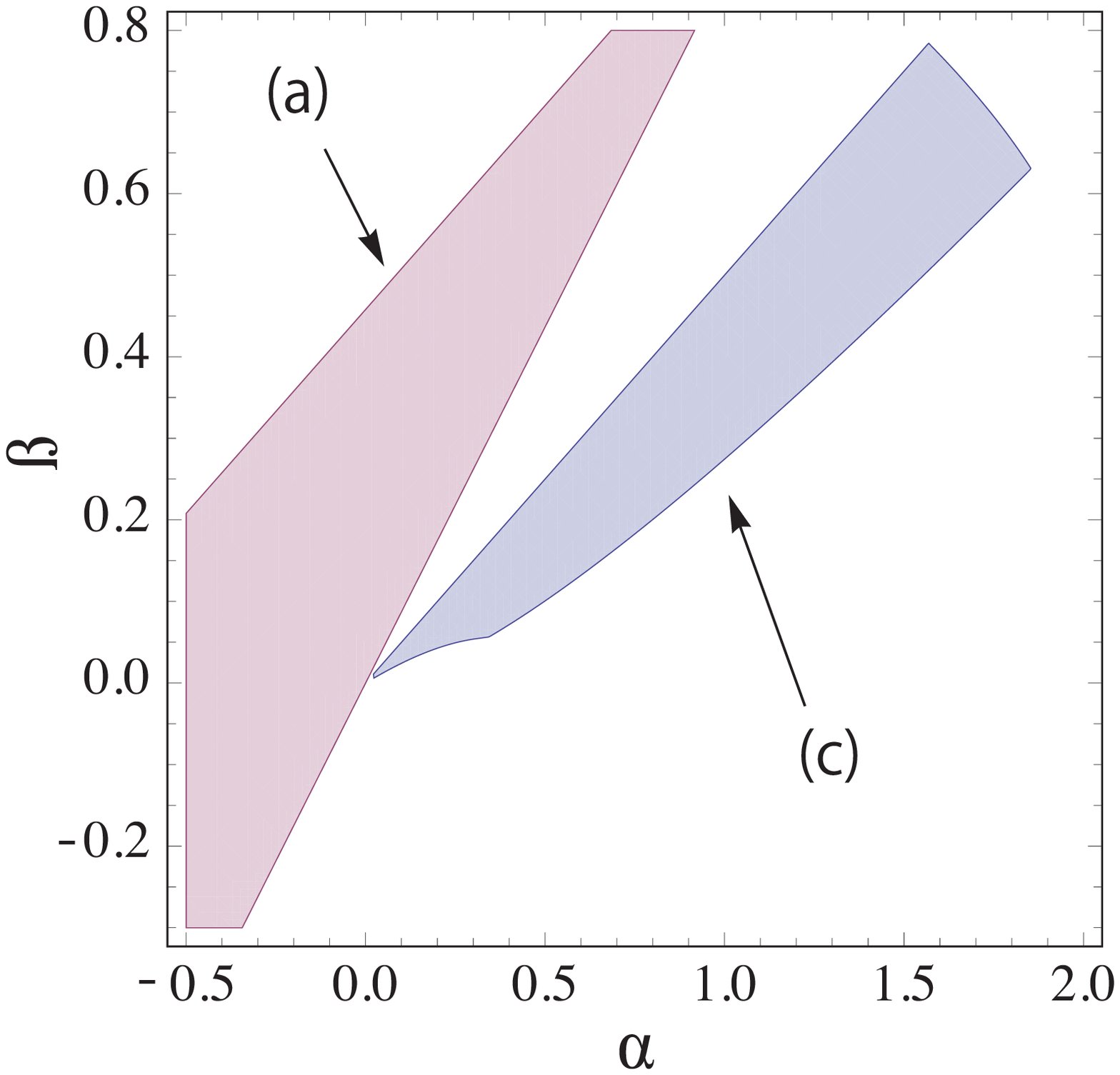}
    \end{center}
   \end{minipage}
   \begin{minipage}{0.5\textwidth}
    \begin{center}
     \includegraphics[scale=0.4]{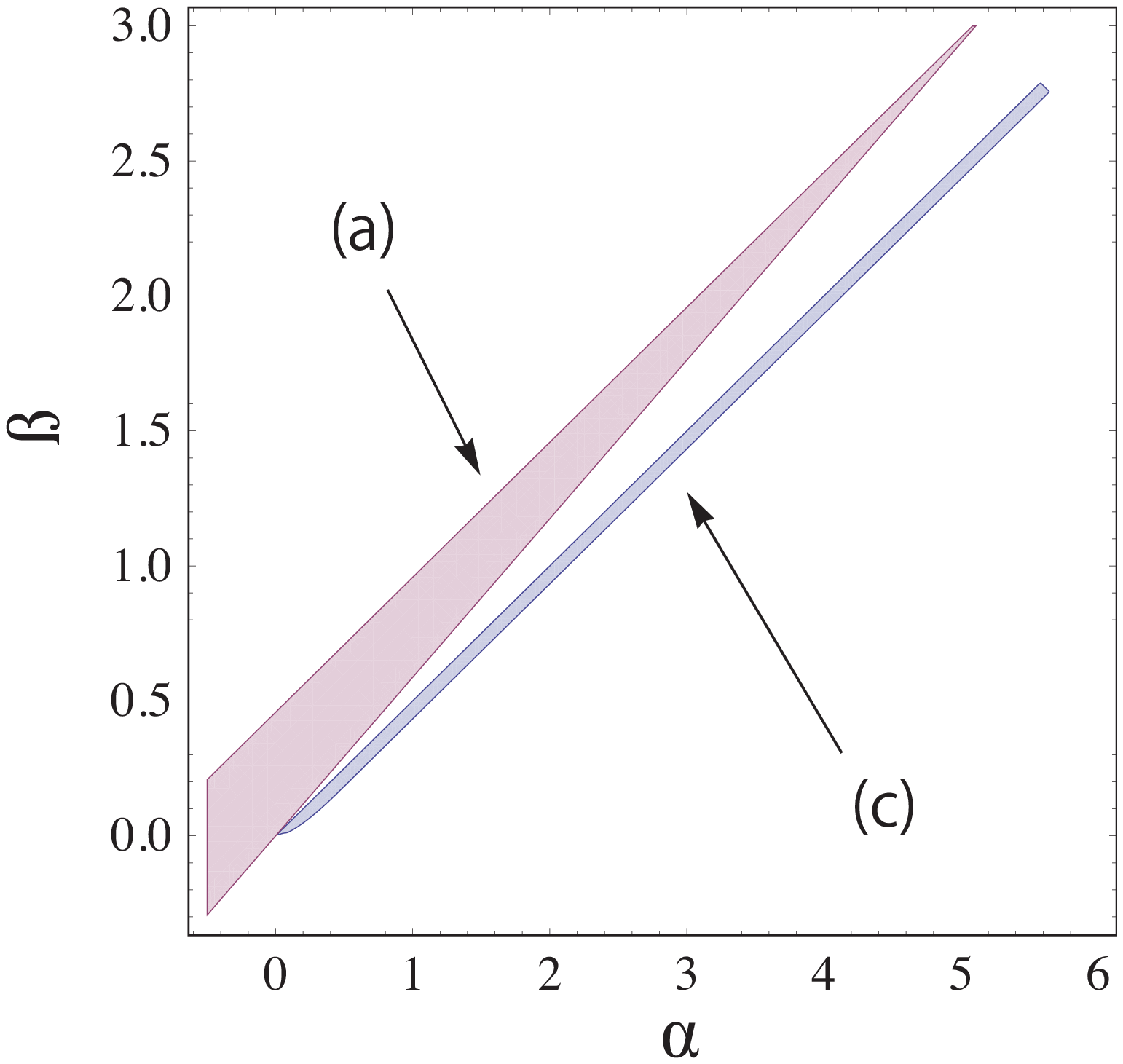}
    \end{center}
   \end{minipage}
  \end{tabular}
  \caption{The allowed parameter space which satisfies the constraint (a)
from the gravitational Cherenkov radiation (\ref{constraint}) and the 
constraint (c) from (\ref{condition:1}), (\ref{condition:2}), and 
(\ref{condition:3}). 
The left panel assumes $p=1$ and $q=1/2$, while
the right panel does $p=1$ and $q=5/2$. 
  }
  \label{fig:parameter1}
\end{figure}

\begin{figure}[t]
  \begin{tabular}{cc}
   \begin{minipage}{0.5\textwidth}
    \begin{center}
     \includegraphics[scale=0.4]{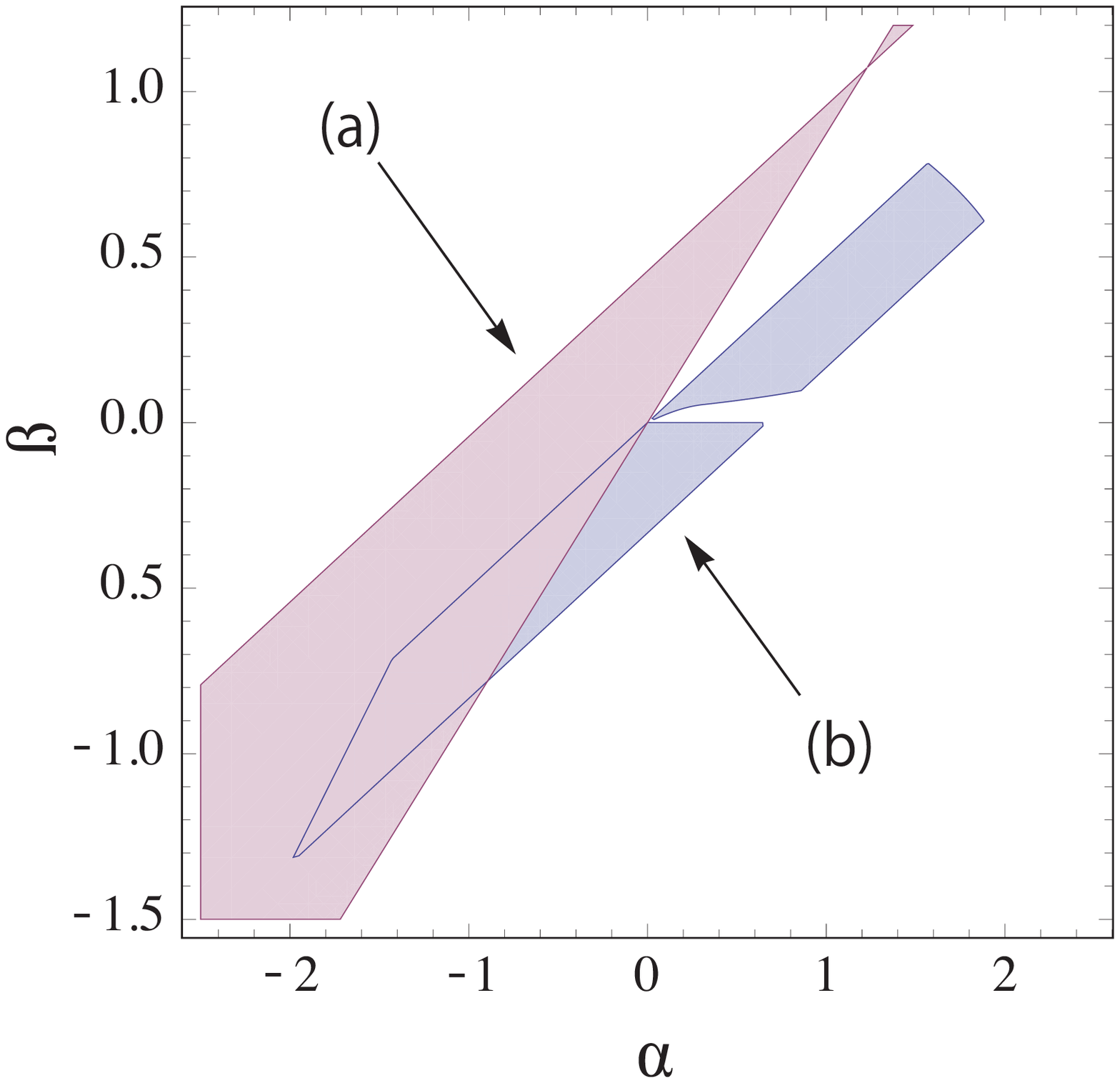}
    \end{center}
   \end{minipage}
   \begin{minipage}{0.5\textwidth}
    \begin{center}
     \includegraphics[scale=0.4]{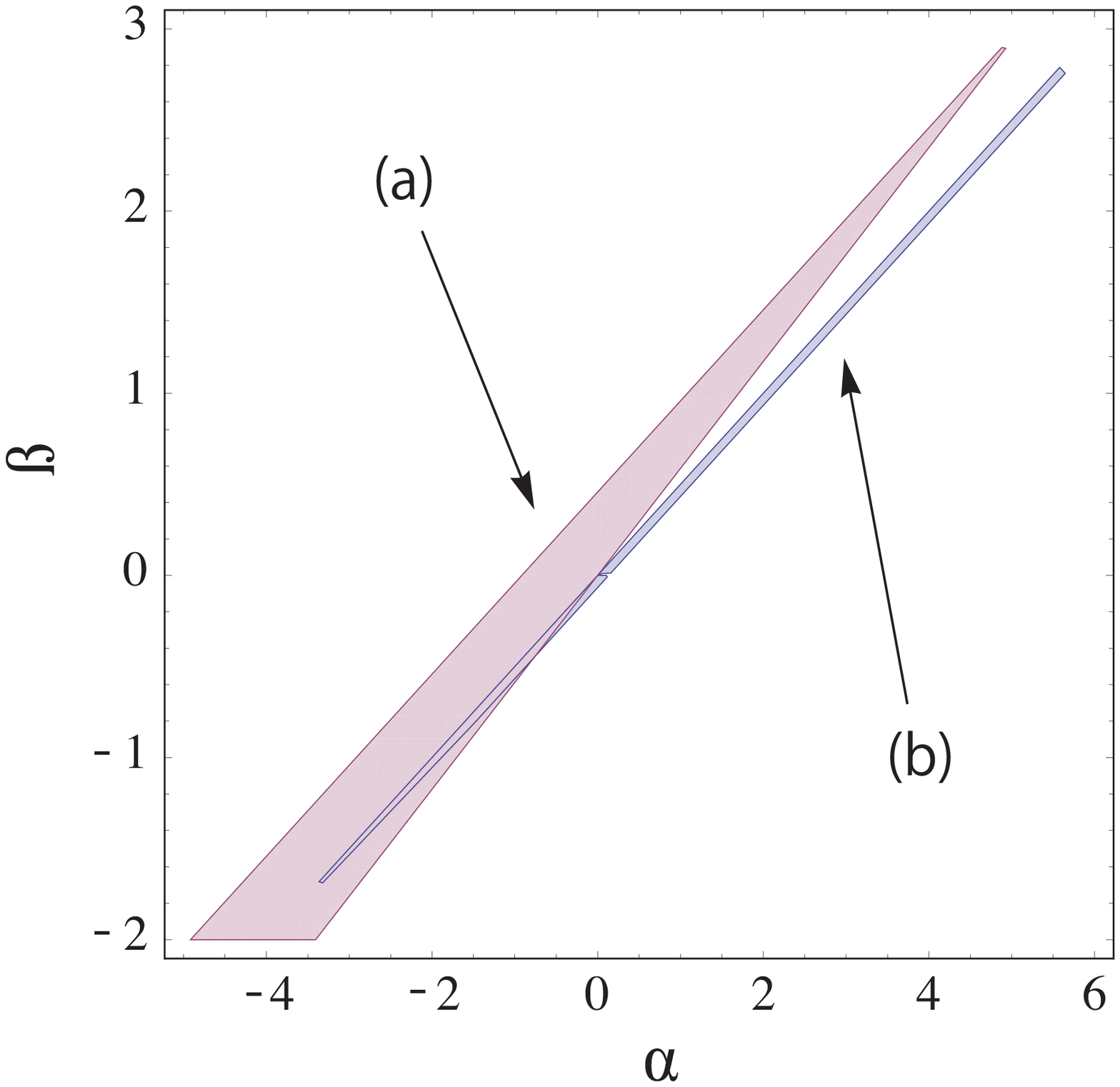}
    \end{center}
   \end{minipage}
  \end{tabular}
  \caption{The allowed parameter space which satisfies the constraint (a)
from the gravitational Cherenkov radiation (\ref{constraint}) and the 
constraint (b) from (\ref{condition:1}).
The left panel assumes $p=1$ and $q=1/2$, while
the right panel does $p=1$ and $q=5/2$. 
  }
  \label{fig:parameter2}
\end{figure}

Figure~\ref{fig:parameter1} shows the allowed regions to satisfy 
the constraint (a) and the constraint (c) for $p=1$ and 
$q=1/2$ (left panel) and $p=1$  and $5/2$ (right panel), where we adopt 
$\Omega_{m0}h^2=0.1344$  and $\Omega_{r0}h^2=4.17\times 10^{-5}$ with 
$h=0.7$. In this case, we see that there is no overlap region 
except for $\alpha=0$ and $\beta = 0$. Thus, the constraint 
from the gravitational Cherenkov radiation is crucial.
Figure~\ref{fig:parameter2} is the same as figure~\ref{fig:parameter1}, 
but for the constraint (a) and the constraint (b). 
We see that the allowed region in parameter space is significantly reduced, 
by combining with the constraint from gravitational Cherenkov radiation (a). 
Especially, there is no overlap region with the positive 
values of $\alpha$ and $\beta$, in figure~\ref{fig:parameter2}.
In general one can show that both the constraints (a) and (b) impose 
$\alpha$ and $\beta$ to be negative or zero  
 for any values of $p\geq 1$ and $q \geq 0$ 
(see appendix \ref{App:negativeab}).

We must further include the constraint from cosmological 
observations (d). 
The authors in ref.~\cite{CovGalileon2} 
investigated the constraint on the covariant galileon model 
($p=1$ and $q=1/2$) from the observational data of 
type Ia supernovae, the shift parameter from the cosmic 
microwave background, and the baryon acoustic oscillations.
They showed the early tracking solution,
corresponding to the case of (b),
is disfavored by the cosmological constraint (d).
On the other hand the solutions that approach the tracker 
solution only at late times, corresponding to the case of (c),
are favored (also see \cite{ObsExtGalileon})
taking small spatial curvature into account. 
However, the latter case is significantly constrained by combining 
the constraint (a), though we do not take the spatial curvature
into account.

Thus, the constraint from the gravitational Cherenkov radiation
plays a very important role to reduce the allowed 
parameter-space of the extended galileon model. In ref.~\cite{ISW}, it 
is demonstrated that the integrated Sachs Wolfe effect 
derives a stringent constraint on a subclass of the galileon model. 
Further tight constraint could be obtained by combining these 
constraints.

\section{Conclusion}
\label{sec:5}
In this paper, we studied constraints on the general scalar-tensor theories
on a cosmological background, 
whose propagation speed of gravitational waves
differs from the speed of light, 
using the survival of high energy cosmic ray against the gravitational Cherenkov radiation. 
In these theories, the coupling of the scalar field $\phi$ and its kinetic term $X$
with gravity causes the violation of Lorentz invariance in a cosmological 
background, leading to a time-dependent propagation speed of gravitational
waves.
We demonstrated that such a model can be constrained using the survival of high 
energy cosmic ray against the gravitational Cherenkov radiation. 

We first considered constraints on the purely kinetic coupled gravity
and found that the conditions for the existence of a desired late-time solution 
and avoiding ghost-instability is $0<\delta<2/5$ 
while the constraint from the gravitational Cherenkov radiation gives 
$\delta > 1-{\cal O}(\epsilon)$, where $\epsilon=2\times 10^{-15}$.
Thus the purely kinetic coupled gravity is inconsistent 
with the argument of the gravitational Cherenkov radiation.

We also focused our investigation on the extended galileon model,
which is a generalization of the covariant galileon model
in the framework of the most general second-order scalar-tensor theory.
We showed that there is no allowed parameter space 
except for $\alpha=\beta=0$ by combining 
the condition for avoiding ghost-instabilities and 
the constraints from the gravitational Cherenkov radiation
if the initial condition of $r_1$ is sufficiently small.
Even if the initial condition of $r_1$ is placed near the 
tracker $r_1=1$, the allowed parameter space is tightly 
constrained by combining the gravitational Cherenkov radiation
and cosmological constraint such as type Ia supernovae, the shift 
parameter from cosmic microwave background,
baryon acoustic oscillations.

Thus the constraint from the gravitational Cherenkov radiation 
is important to constrain the general second-order scalar-tensor theories
on a cosmological background, 
whose propagation speed of gravitational waves is less than the speed of light.

\acknowledgments 
This work was
supported in part by JSPS Grant-in-Aid for Scientific Research
No.~21540270 and No.~21244033 and JSPS Core-to-Core Program
``International Research Network for Dark Energy''.
R.K. acknowledges support by a research assistant program
of Hiroshima University.
R.K. was also supported in part by a Grant-in-Aid for JSPS Fellows.
K.Y. thanks M. Yamaguchi for a useful discussion at a workshop held
in Takehara on the sound speed of the tensor mode 
in the most general second-order scalar-tensor theory.
We also thank T. Kobayashi for useful discussions, 
when the authors initiated this work.

\appendix
\section{Scalar perturbations}
\label{App:Coefficients1}
Here we summarize the scalar perturbations 
including coefficients in the most general second-order
scalar-tensor theory derived in \cite{G-inf}.
For the unitary gauge $\phi=\phi(t)$ with the line element 
$ds^2=-N^2dt^2+\gamma_{ij}(dx^i+N^idt)(dx^j+N^jdt)$,
where $N=1+\alpha$, $N_i=\partial_i\beta$, 
and $\gamma_{ij}=a^2(t)e^{2\zeta}\delta_{ij}$,
the quadratic action for the scalar perturbations
after solving the constraint equations can be written as
\begin{eqnarray}
  S_S^{(2)}=\int dtd^3x a^3 
  \left[{\cal G}_S {\dot \zeta}^2 -\frac{{\cal F}_S}{a^2}({\vec \nabla}\zeta)^2\right],
\end{eqnarray}
where
\begin{eqnarray}
{\cal F}_S&\equiv&\frac{1}{a}\frac{d}{dt}\left(\frac{a}{\Theta}{\cal G}_T^2\right)-{\cal F}_T,
\label{Fs}\\
{\cal G}_S&\equiv&\frac{\Sigma}{\Theta^2}{\cal G}_T^2+3{\cal G}_T,
\label{Gs}
\end{eqnarray}
and 
\begin{eqnarray}
\Theta&\equiv&-\dot\phi XG_{3X}+
2HG_4-8HXG_{4X}
-8HX^2G_{4XX}+\dot\phi G_{4\phi}+2X\dot\phi G_{4\phi X}
\nonumber\\&&
-H^2\dot\phi\left(5XG_{5X}+2X^2G_{5XX}\right)
+2HX\left(3G_{5\phi}+2XG_{5\phi X}\right),
\\
\Sigma&\equiv&X K_X+2X^2K_{XX}+12H\dot\phi X G_{3X}
+6H\dot\phi X^2 G_{3XX}-2X G_{3\phi} -2 X^2 G_{3\phi X}\nonumber\\
&&-6H^2G_4 +6 \bigr[H^2(7XG_{4X}+16X^2G_{4XX}+4X^3G_{4XXX})\nonumber\\
&&-H \dot\phi (G_{4\phi}+5XG_{4\phi X}+2X^2G_{4\phi XX})\bigl]
+30H^3 \dot\phi X G_{5X}+26H^3 \dot\phi X^2 G_{5XX}\nonumber\\
&&+4H^3 \dot\phi X^3 G_{5XXX}-6H^2 X (6G_{5\phi}+9XG_{5\phi X}+2X^2G_{5\phi XX}).
\end{eqnarray}
The propagation speed of the scalar perturbations is defined as 
\begin{eqnarray}
c_S^2&\equiv&\frac{{\cal F}_S}{{\cal G}_S}.
\label{cs2}
\end{eqnarray}

\section{Coefficients and propagation speed in various regimes}
\label{App:Coefficients2}
In this appendix, we summarize the coefficients 
and propagation speed in the tensor and scalar
perturbation equations in the extended galileon model in various regimes, 
derived in \cite{Tsujikawa11}.

In the regime, $r_1=1$ and $r_2 \ll 1$, 
the coefficients (\ref{Gt}), (\ref{ct2}), (\ref{Gs}), 
and (\ref{cs2}) are given by 
\begin{eqnarray}
  {\cal G}_S|_{r_1=1, r_2 \ll 1} & \simeq & 6q\left[p-3(\alpha-2\beta)q\right]r_{2},\\
  c_S^2|_{r_1=1, r_2 \ll 1} & \simeq &\bigl\{4p^{3}(\Omega_{r}+3)-2p^{2}\{(\Omega_{r}+3)(6\beta-3\alpha+2)\nonumber\\
  &&-2q[3\Omega_{r}+11-3(\alpha-2\beta)(\Omega_{r}+3)]\}-3\{\beta(\Omega_{r}+3)\nonumber \\
  && +8q^{3}(\Omega_{r}+5)(\alpha-2\beta)-2q^{2}(7\Omega_{r}+27)(\alpha-2\beta)\nonumber\\
  &&+q[3\alpha(\Omega_{r}+3)-2\beta(5\Omega_{r}+17)]\}\nonumber \\
  && -p\{(\Omega_{r}+3)(3\alpha-12\beta-1)+4q^{2}[(\alpha-2\beta)(9\Omega_{r}+33)-2(\Omega_{r}+5)]\nonumber \\
  && +q[12(2\beta-\alpha)(3\Omega_{r}+10)+6\Omega_{r}+22]\}\bigl\}\nonumber \\
  &&\times 1/[24q^{2}(2p+4q-1)\{p-3(\alpha-2\beta)q\}],\\
  {\cal G}_T|_{r_1=1, r_2 \ll 1} & \simeq &\frac{1}{2}\left[2+3(\alpha-2\beta)r_{2}\right],\\
  c_T^2|_{r_1=1, r_2 \ll 1} & \simeq &1-\{6[2(\alpha-2\beta)q+3\beta]p+24(\alpha-2\beta)q^{2}\nonumber \\
    &&~~~~~~+3\beta(16q-3)+3\beta(2p+4q-1)\Omega_{r}\}/{4q(2p+4q-1)}r_{2}.
\end{eqnarray}
At the de Sitter point, $r_1=r_2=1$, 
the coefficients (\ref{Gt}), (\ref{ct2}), (\ref{Gs}), 
and (\ref{cs2}) are given by 
\begin{eqnarray}
  {\cal G}_S|_{\rm de~Sitter} &=&\frac{6(p+2q)(3\alpha-6\beta+2)[p-3(\alpha-2\beta)q]}
  {[2p-6(\alpha-2\beta)q-3\alpha+6\beta-2]^{2}},\\
  c_S^2|_{\rm de~Sitter} &=&\{6\beta+4p^{2}+p\,[9(\alpha-2\beta)^{2}+3\alpha
  -12\beta+4q(6\beta-3\alpha+2)-2]\nonumber \\
  &&+3(\alpha-2\beta)[3\beta+q(9\alpha-12\beta-8q+6)]\}\nonumber \\
  && {}\times\frac{3(2\beta-\alpha)(2q+1)+2p-2}{6(6\beta-3\alpha-2)(p+2q)(2p+4q-1)(p-3\alpha q+6\beta q)},\\
  {\cal G}_T|_{\rm de~Sitter} &=&\frac{1}{2}\left(3\alpha-6\beta+2\right),\\
  c_T^2|_{\rm de~Sitter} &=&\frac{2(2p+4q-1)-3\alpha}{(2p+4q-1)(3\alpha-6\beta+2)}.
\end{eqnarray}
In the regime, $r_1 \ll 1$ and $r_2 \ll 1$, 
the coefficients (\ref{Gt}), (\ref{ct2}), (\ref{Gs}), 
and (\ref{cs2}) are given by 
\begin{eqnarray}
  {\cal G}_S|_{r_1 \ll 1, r_2 \ll 1}& \simeq &3(p+3q)(2p+6q-1)\beta r_{1}^{(p-1)/(2q+1)}r_{2},\\
  c_S^2|_{r_1 \ll 1, r_2 \ll 1}& \simeq &\frac{p+3q-2}{2(p+3q)(2p+6q-1)}\left(1+\Omega_{r}\right),\\
  {\cal G}_T|_{r_1 \ll 1, r_2 \ll 1}& \simeq &1-3\beta r_{2}r_{1}^{(p-1)/(2q+1)},\\
  c_T^2|_{r_1 \ll 1, r_2 \ll 1}& \simeq &1+\frac{3(4p+12q-5-3\Omega_{r})}{4p+12q-2}\beta r_{1}^{(p-1)/(2q+1)}r_{2}.
\end{eqnarray}

\section{Other fixed points}
\label{App:otherfixedpoint}
There also exist the other fixed points found by De Felice and Tsujikawa \cite{Tsujikawa11},
which is characterized by the equation,
\begin{equation}
p(3\alpha-4\beta+2)r_{i}^{2}+[2\beta(p+3q)-3\alpha(p+2q)]r_{i}+2\beta(p+3q)=0,
\end{equation}
where $r_i=r_a$ and $r_b$, and 
\begin{eqnarray}
  r_{a,b}&=&\frac{3\alpha(p+2q)-2\beta(p+3q)\pm \sqrt{\Delta}}{2p(3\alpha-4\beta+2)},\\
  \Delta&=&[2\beta(p+3q)-3\alpha(p+2q)]^2-8\beta p (3\alpha-4\beta+2)(p+3q).
\end{eqnarray}

\section{Constraint on the values of $\alpha$ and $\beta$}
\label{App:negativeab}
In this appendix, we assume $p\geq 1$, $q \geq 0$, and $r_2=1-\Omega_{m0}-\Omega_{r0}$, 
where $\Omega_{m0}h^2=0.1344$  and $\Omega_{r0}h^2=4.17\times 10^{-5}$ with 
$h=0.7$. One can show that $\alpha$ and $\beta$ must be negative or zero to 
satisfy both the constraints (a) and (b) for any values of
$p\geq 1$ and $q \geq 0$. This can be proved as follows.
The upper bound of the constraint (b) is 
determined by the straight line in the plane of $\alpha$ and $\beta$,
\begin{eqnarray}
  \beta=\frac{1}{2}\alpha-\frac{p-1}{3(2q+1)},
\label{line:cs0}
\end{eqnarray}
which comes from $c_S^2|_{\rm de~Sitter} \geq 0$ (see e.g., the left panel of 
figure~\ref{fig:parameter2}).
On the other hand, the constraint (a) is characterized by the
two straight lines,
\begin{eqnarray}
  \beta={1 \over 2}\alpha + {1 \over 3r_2},
\label{gcr:line1}
\end{eqnarray}
and
\begin{eqnarray}
  \beta=\frac{2(p+2q)(2q+pr_2)}{(2p+4q-1)(4q-3+3r_2+2pr_2-\Omega_r)}\alpha.
\label{gcr:line2}
\end{eqnarray}
Note that the lines of eqs.~(\ref{gcr:line1}) and (\ref{line:cs0})
are parallel each other. Since we have
\begin{eqnarray}
\frac{2(p+2q)(2q+pr_2)}{(2p+4q-1)(4q-3+3r_2+2pr_2-\Omega_r)}>{1\over 2},
\label{slope}
\end{eqnarray}
at present epoch, the lines of eqs.~(\ref{gcr:line1}) and (\ref{gcr:line2}) 
intersect at a point with $\alpha>0$ and $\beta>0$. 
(\ref{slope}) means the slope of the line 
(\ref{gcr:line2}) is larger than that of the line (\ref{line:cs0}), 
then the lines of eqs.~(\ref{line:cs0}) and (\ref{gcr:line2})
intersect at a point with $\alpha\leq0$ and $\beta\leq0$, 
Therefore, $\alpha$ and $\beta$ must be negative or zero to satisfy the 
constraints (a) and (b) in the case of $p \geq 1$ and $q \geq 0$.


\end{document}